\documentclass[aip, superscriptaddress, preprint]{revtex4}
\usepackage{graphicx}% Include figure files
\usepackage{dcolumn}% Align table columns on decimal point
\usepackage{bm}% bold math

\begin{document}

\title{Room-temperature generation of giant pure spin currents 
\\using Co$_{2}$FeSi spin injectors}% Force line breaks with \\

\author{Takashi Kimura}
\email[]{kimura@ifrc.kyushu-u.ac.jp }
\affiliation{Advanced Electronics Research Division, 
INAMORI Frontier Research Center, Kyushu University, 744 Motooka, Fukuoka, 819-0395, Japan}
\affiliation{CREST, Japan Science and Technology Agency, Sanbancho, Tokyo 102-0075, Japan}

\author{Naoki Hashimoto}
\affiliation{Department of Electronics, Kyushu University, 744 Motooka, Fukuoka 819-0395, Japan}

\author{\\ Shinya Yamada}
\affiliation{Department of Electronics, Kyushu University, 744 Motooka, Fukuoka 819-0395, Japan}

\author{Masanobu Miyao}
\affiliation{Department of Electronics, Kyushu University, 744 Motooka, Fukuoka 819-0395, Japan}
\affiliation{CREST, Japan Science and Technology Agency, Sanbancho, Tokyo 102-0075, Japan}

\author{Kohei Hamaya}
\email[]{hamaya@ed.kyushu-u.ac.jp}
\affiliation{Department of Electronics, Kyushu University, 744 Motooka, Fukuoka 819-0395, Japan}
\affiliation{PRESTO, Japan Science and Technology Agency, Sanbancho, Tokyo 102-0075, Japan}

%\affiliation{
%%Second institution and/or address\\

%\begin{abstract}
%\end{abstract}
\date{\today}% It is always \today, today,

%\end{abstract}
%\pacs{72.25.Dc, 72.25.Hg, 72.25.Mk}% PACS, the Physics and Astronomy
\maketitle
%\section{INTRODUCTION}

{\bf 
Generation, manipulation, and detection of a pure spin current, i.e., the flow of spin angular momentum without a charge current\cite{Jedema01-1, Jedema01-2, Chappert, Zutic, Saitoh, Kimura02}, are prospective approaches for realizing next-generation spintronic devices with ultra low electric power consumptions. Conventional ferromagnetic electrodes such as Co and NiFe have so far been utilized as a spin injector for generating the pure spin currents in nonmagnetic channels\cite{Jedema01-1, Jedema01-2,  Kimura02, Kimura03, Kimura04, Yang, Kimura05, Tinkham, Johnson, Mihajlovic, idzuchi}. However, the generation efficiency of the pure spin currents is extremely low at room temperature, giving rise to a serious obstacle for device applications. Here, we demonstrate the generation of giant pure spin currents at room temperature in lateral spin valve devices with a highly ordered Heusler-compound Co$_2$FeSi spin injector. The generation efficiency of the pure spin currents for the Co$_2$FeSi spin injectors reaches approximately one hundred times as large as that for NiFe ones
, indicating that Heusler-compound spin injectors enable us to materialize a high-performance lateral spin device. The present study is a technological jump in spintronics and indicates the great potential of ferromagnetic Heusler compounds with half metallicity for generating pure spin currents.
 }

\newpage

Electrical spin injection from a ferromagnet (F) into a nonmagnet (N) can generate a spin current, i.e., the flow of spin angular momentum, even in a nonmagnetic channel\cite{Johnson1}.   In general, the spin current is induced by diffusing non-equilibrium spin accumulations in the vicinity of the F/N interface under the spin injection. However, since the difference in the density of states between majority and minority spins, i.e., spin polarization $P$, is not so large for a conventional F such as Co or NiFe (Py), the induced spin current in the N mainly returns back to the F (Fig.\ 1a). This gives rise to an extremely low injection efficiency of the spin current in the N\cite{Kimura02, van Son, Schmidt}. If we utilize a perfectly spin-polarized F, so called a half-metallic ferromagnet (HMF)\cite{Groot}, as a spin injector, fully spin-polarized electrons can be injected into the N and the backflow of the spin currents can be completely suppressed, resulting in a dramatical improvement of the injection efficiency of the spin currents in the N (Fig.\ 1b). Also, one can extract a charge current by using nonlocal electrical spin injection in a mesoscopic lateral geometry, and can transfer only a spin current without the charge current, i.e., a pure spin current (Fig.\ 1c), in the nonmagnetic channel. In this scheme, using HMF spin injectors is a key for generating a {\it giant} pure spin current in the N (Fig.\ 1d).

As materials with half metallicity, we focus on Co-based Heusler compounds which enable huge tunnel magnetoresistance (TMR) and giant magnetoresistance (GMR) effects in vertical stacking device structures\cite{Inomata, Felser, Sakuraba1}. Despite these high performances, none of the lateral spin transports using the Heusler-compound electrodes have been reported yet. Thus, the combination of the high-performance Co-based Heusler compounds with laterally configured device structures is a prospective challenge for highly efficient generation of the pure spin currents. In this work, we show that a Co-based Heusler compound, Co$_2$FeSi (CFS), enables the highly efficient injection of the spin currents.

Our device structure is a lateral spin valve (LSV) consisting of the CFS spin injector and detector bridged by a Cu strip (Fig.\ 2a), where the CFS thin film with highly ordered $L2_{1}$ structures has been epitaxially grown on Si(111)\cite{Yamada}. Details of the growth of the CFS thin films and the fabrication processes of the LSVs are given in the Methods sections. As shown in Fig.\ 2b, a pure spin current generated by the nonlocal spin injection from CFS1 can be detected by CFS2 after the propagation of 600-nm distance in the Cu strip. 
Figure\ 2c shows a nonlocal magnetoresistance of the CFS/Cu LSV measured at room temperature (RT), together with that of a Py/Cu LSV. Here, the size of the CFS/Cu junction is three times as large as that of the Py/Cu junction. Note that a giant spin signal ($\Delta R_{\rm S}$) of 2.3 m$\Omega$ is seen for the CFS/Cu LSV (Fig.\ 2c), which is approximately ten times as large as that for the Py/Cu LSV. Since the spin injection efficiency is inversely proportional to the size of the F/N junctions \cite{Kimura02}, the giant $\Delta R_{\rm S}$ demonstrated in the CFS/Cu LSV with larger sizes in the junctions implies a great possibility of the present CFS/Cu LSV.  We also show the local spin valve signal of 4.5 m$\Omega$ at RT for the same CFS/Cu LSV (Fig.\ 2d). The value of 4.5 m$\Omega$ is almost twice of the non-local $\Delta R_{\rm S}$, in reasonable agreement with the previous reports \cite{Jedema01-2, Kimura02}.   
This means that one dimensional spin diffusion model well describes 
the spin transport in the present CFS/Cu LSV.  
Figure 2e shows a dependence of $\Delta R_{\rm S}$ 
on the bias current density ($J_{\rm inj}$) at the injecting junction for the CFS/Cu LSV, which is almost same as that for the Py/Cu LSV.  The reduction of the $\Delta R_{\rm S}$ is less than 20 \% even under a high bias current density ($\sim$ 10$^{11}$ A/m$^2$), indicating much superior property compared to the LSV consisting of the high resistive tunnel junctions, where the $\Delta R_{\rm S}$ drastically decreases even at low bias current density ($\sim 10^8$ A/m$^2$)\cite{Tinkham}. The temperature dependence of $\Delta R_{\rm S}$ for the CFS/Cu LSV is also almost same as that for the Py/Cu LSV, where the $\Delta R_{\rm S}$ takes a maximum value around 20 K, below which the $\Delta R_{\rm S}$ decreases with decreasing temperature (Fig.\ 2f).  This behavior can be explained by an enhancement in the spin-flip scattering at the Cu surface for the CFS/Cu LSV below 20 K, as discussed in Ref \cite{Kimura05}.  Surprisingly, the $\Delta R_{\rm S}$ for the CFS/Cu exceeds 10 m$\Omega$ below 70 K (inset of Fig.\ 2f).  From these results, we recognize that the present CFS/Cu LSV can be treated as conventional ohmic LSVs and can generate a giant pure spin current with a much less electric power than previously reported LSVs\cite{Jedema01-1, Jedema01-2, Kimura02, Kimura03, Kimura04, Yang, Kimura05, Tinkham, Johnson, Mihajlovic, idzuchi}.

To quantitatively evaluate the device performance of the present LSVs from the nonlocal spin signals, we measured $\Delta R_{\rm S}$ of CFS/Cu LSV devices with various distances ($d$), together with Py/Cu LSV devices as references, where $d$ is the centre-centre distance between spin injector and detector. Here, we introduce a characteristic value in the LSV devices by extending the resistance change area product, commonly utilized to characterize the device performances in the vertical spin devices\cite{Nakatani2, Sakuraba2}. The resistance change area product for the nonlocal spin signal, i.e., $\Delta R_{\rm S} A$, is defined as $\Delta R_{\rm S} ( S_{\rm inj} S_{\rm det} / S_{\rm N} )$, where $S_{\rm inj}$, $S_{\rm det}$, and $S_{\rm N}$ are 
the junction sizes in the spin injector and detector, and the cross section of the nonmagnetic strip. This $\Delta R_{\rm S} A$  allows us to equivalently compare the generation efficiency of the pure spin current between our CFS/Cu and the other conventional F/N LSV devices. The plot of $\Delta R_{\rm S} A$ versus $d$ at RT for CFS/Cu LSVs and Py/Cu LSVs is shown in Fig. 3, together with that at 80 K in the inset. The $\Delta R_{\rm S} A$ is increased with decreasing $d$ for both series of the CFS/Cu and Py/Cu LSV devices. By solving one dimensional spin diffusion equation\cite{Takahashi, Kimura02} (see Supplementary information), $\Delta R_{\rm S} A$ can be expressed as 
\begin{equation}
\Delta R_{\rm S} A \approx
\frac{ \left( \frac{ P_F}{(1-P_F^2)} \rho_{\rm F} \lambda_{\rm F} 
+ \frac{P_I}{(1-P_I^2)} {\rm RA}_{\rm F/N} \right)^2 }
{\rho_{\rm N} \lambda_{\rm N} \sinh \left( {d}/{\lambda_{\rm N}} \right)}, 
\end{equation}
where $P_{\rm F}$ and $P_{\rm I}$ are, respectively, the bulk and interface spin polarizations for F,  $\lambda_{\rm F}$ and $\lambda_{\rm N}$ are the spin diffusion lengths for F and N, and, $\rho_{\rm F}$ and $\rho_{\rm N}$ are the resistivities for F and N, respectively. 
As shown in Fig.\ 3, the plots of $\Delta R_{\rm S} A$ versus $d$ for both series of the CFS/Cu and Py/Cu LSV devices are well reproduced by the fitting curves with $\lambda_{\rm Cu} = $ 500 nm and $\lambda_{\rm Cu} = $ 1300 nm at RT and 80 K\cite{Jedema01-1, Jedema01-2, Kimura02,Kimura03}, respectively. For the Py/Cu LSVs, assuming $\lambda_{\rm Py, RT} =$ 3 nm and $\lambda_{\rm Py, 80 K} =$ 5 nm, we obtained a reasonable $P_{\rm Py}$ of $0.3$ and 0.35 at RT and 80 K, respectively\cite{Jedema01-1, Jedema01-2,Kimura02}. Thus, the above equation is a reliable for expressing the generation efficiency of the pure spin current among various LSVs.  
We then roughly estimate the spin polarization for CFS ($P_{\rm CFS}$).  
Since it is impossible to determine the spin polarization and the spin 
diffusion length independently from the present results, 
we assume that $\lambda_{\rm CFS}$ is the same order of that for CFSA 
 (see Supplementary information).  If we use $\lambda_{\rm CFS, RT} =$ $2 \sim 4$ nm and $\lambda_{\rm CFS, 80 K} =$ $3 \sim 6$ nm, $P_{\rm CFS}$ can be estimated to be 0.56 $\pm$ 0.10 at RT and 0.67 $\pm$ 0.11 at 80 K, similar to the $P$ estimated from the analysis of the current perpendicular GMR effects\cite{Sakuraba2, Nakatani2}.   Although the present CFS epitaxial layers have highly ordered structures with a high magnetic moment above 5 $\mu_{\rm B}$/f.u\cite{Yamada}, the value is still smaller than 6 $\mu_{\rm B}$/f.u. in the perfectly ordered CFS\cite{Inomata, Felser}.  Since further enhancement in $P_{\rm CFS}$ will be achieved by improving the crystal growth technique, a scaling characteristic with $P = 1$ will be obtainable ultimately (see dashed line).  It should be noted that the resistivity for the ferromagnetic electrode ($\rho_{\rm F}$) and the interface resistance area product (RA$_{\rm F/N}$) are also important factors in Eq.\ (1).  Therefore, the relatively large $\rho_{\rm CFS}$ and RA$_{\rm CFS/Cu}$  compared to those in the Py/Cu LSV (see Method) 
 are also advantages for obtaining large $\Delta R_{\rm S} A$.  
%Using this $\Delta R_{\rm S} A$, we compare 
%the device performance of the CFS/Cu LSVs 
%to the Py/Cu LSVs more quantitatively.  
%For equivalent comparisons, we concentrate on $\Delta R_{\rm S} A$ at $d =$ 100 nm. 

Our data for the CFS/Cu LSVs is approximately one hundred times as large as that for the 
Py/Cu LSV, indicating a significant improvement of the generation efficiency of the pure spin current using ohmic junctions.  
The present result is a markedly technological advance in spintronics using pure spin currents, generated by Heusler-compound spin injectors.

\newpage

\begin{figure}[!t]
\includegraphics[width=14.5cm]{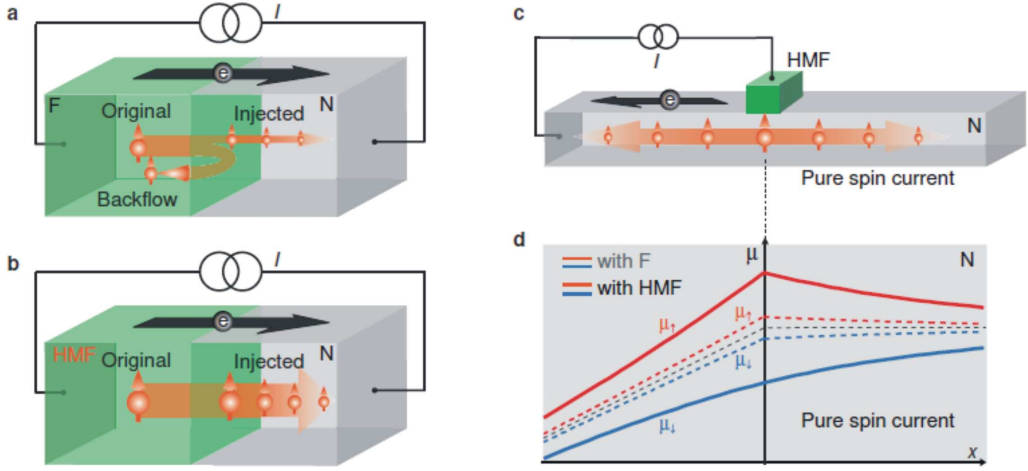}
\caption{
{\bf Concept of efficient generation of a pure spin current.} {\bf a,b,} Schematic diagrams of the electrical spin injection from a conventional ferromagnet (F) or a half-metallic (HM) F into a nonmagnet (N). For the conventional F, most of the original spin currents go back to the F (a backflow of the spin current), giving rise to a significant reduction in the injected spin current. For the HMF, the original spin current is fully injected into the N without the backflow. {\bf c,} Generation of a pure spin current by using nonlocal spin injection. The electron charges are extracted toward left hand side while the spin currents diffuse into both side symmetrically. {\bf d,} Spatial distributions of the spin-dependent electro-chemical potentials, ($\mu_{\uparrow}$, $\mu_{\downarrow}$), in the N. Although the charge current $\propto \partial (\mu_{\uparrow} + \mu_{\downarrow}) / \partial x$ is zero in the right hand side, a finite spin current $\propto  \partial(\mu_{\uparrow} - \mu_{\downarrow}) / \partial x$ is generated over the spin diffusion length. Thus, the pure spin current can be generated in the right hand side of the N.  }  
\end{figure}

\begin{figure}[!t]
\includegraphics[width=14.5cm]{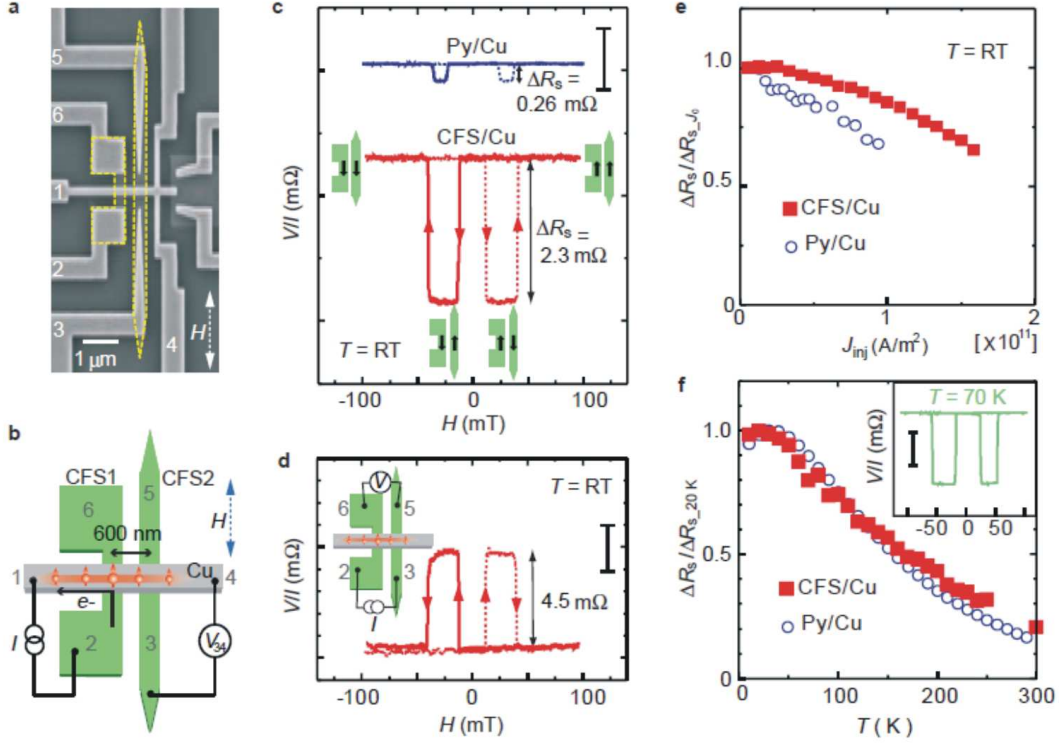}
\caption{{\bf Giant nonlocal spin valve effect.}  {\bf a,} A scanning electron microscope image of the fabricated Co$_{2}$FeSi(CFS)/Cu lateral spin valve. 
{\bf b,} Schematic of a nonlocal spin valve measurement. Spin-polarized electrons are injected from contact 2, and electron charges are extracted from contact 1. A nonlocal voltage is measured between contact 3 and contact 4. {\bf c,} A room-temperature nonlocal spin-valve signal for the CFS/Cu LSV, together with that for the Py/Cu LSV. The signal varies according to the relative magnetization orientation of two wire-shaped CFS electrodes, as shown in the inset illustrations.  
{\bf d,} A room-temperature local spin-valve signal for the CFS/Cu LSV. The inset shows the current-voltage probe configuration, i.e. the current is injected from contact 2 and extracted from contact 3, and the voltage is measured between contact 5 and contact 6. The low and high resistance states correspond to the parallel and anti-parallel magnetization alignments, respectively. The expected magnetization configurations agree with those observed in the nonlocal spin valve signal. {\bf e,} Nonlocal spin signal $\Delta R_{\rm S}$ as a function of $J_{\rm inj}$ at the injecting junction, normalized by $\Delta R_{\rm S}$ at a small bias current density of $ J_0 \sim 10^9$ A/m$^2$, for the CFS/Cu LSV (red solid squares) and the Py/Cu LSV (blue open circles).   
%$\Delta R_{\rm S}$ as a function of $J_{\rm inj}$ normalized by $\Delta R_{\rm S}$ at $J_{\rm inj} = 10^7$ A/m$^2$ for CoFe/Al LSV [Ref.\ 12]  is also plotted.  (black open triangles).  
{\bf f,} Temperature dependence of $\Delta R_{\rm S}$ for the CFS/Cu LSV (red solid squares) and the Py/Cu LSV (blue open circles), normalized by $\Delta R_{\rm S}$ at 20 K. The inset shows a nonlocal spin-valve effect of the CFS/Cu LSV at $T = 70$ K. The scale bars in c, d, and the inset of f are 1, 2, and 5 m$\Omega$, respectively.}

\end{figure} 

\begin{figure}[!t]
\includegraphics[width=10cm]{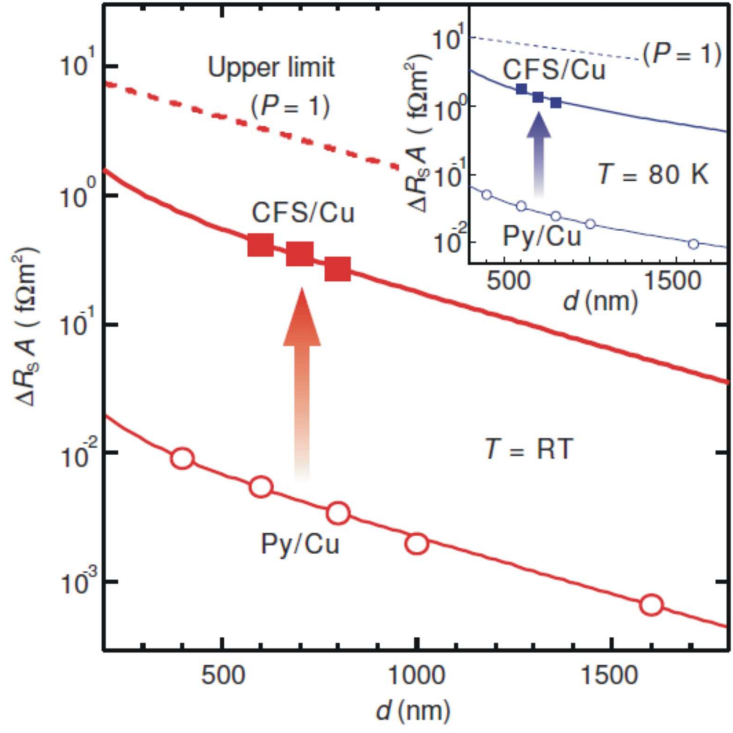}
\caption{{\bf Scaling plot for lateral spin valve devices with metallic junctions.} The resistance change area product for the nonlocal spin signal ($\Delta R_{\rm S} A$) as a function of $d$ for CFS/Cu LSVs (filled squares), together with that for Py/Cu LSVs (open circles). The main panel and inset show the data at RT and 80 K, respectively. The solid curves are fitting results with $\lambda_{\rm Cu} = 500$ nm at RT (red line) and $\lambda_{\rm Cu} = 1300$ nm at 80 K (blue line), and the dashed curves are theoretical upper limits using $P =$ 1, calculated from the equation, 
$\Delta R_{\rm S} A = P^2 (S_{\rm Inj} S_{\rm Det}/S_{\rm N}) \rho_{\rm Cu} \lambda_{\rm Cu} e^{-d/\lambda_{\rm Cu}}$.
}
\end{figure}

%\begin{figure}[!t]
%\includegraphics[width=10cm]{Fig4.eps}
%\caption{ {\bf Trend of spin generation efficiency.} 
%The spin resistance change area product ($\Delta R_{\rm S} A$) for various 
%combinations of a ferromagnetic metal and nonmagnetic one 
%in LSV devices at room temperature. The $\Delta R_{\rm S} A$ corresponds to the spin signal normalized by the geometrical factors (injection and detection junction size, the cross section of the nonmagnetic wire and the separation distance) and the magnitude of the spin signal at $d = 100$ nm.  
%For Ref.\ 2, we assumed that 
%the effective size of junctions for both the spin injector and detector are 150 nm $\times$ 100 nm.  For Ref.\ 8, we assumed that the sizes of the injecting and detecting junctions are 80 nm $\times$ 190 nm and  100 nm $\times$ 190 nm, respectively.  
%For Ref.\ 13, the spin diffusion length of Ag is assumed to be 500 nm.
%}  
%\end{figure}

\clearpage
\section*{Method}
As a spin injector- and detector-material, 25-nm-thick Co$_{2}$FeSi (CFS) films were grown on Si(111) templates by low temperature molecular beam epitaxy\cite{Yamada}. 
Prior to the growth, surface cleaning of substrates was performed with an aqueous HF solution (HF : H$_{2}$O = 1 : 40), and then, they were heat-treated at 450 $^{\circ}$C for 20 min in an MBE chamber with a base pressure of 2 $\times$ 10$^{-9}$ Torr. After the reduction in the substrate temperature down to 100 $^{\circ}$C, we co-evaporated Co, Fe, and Si with stoichiometric chemical compositions by using  Knudsen cells. During the growth, two-dimensional epitaxial growth was confirmed by observing reflection high energy electron diffraction patterns. The formed epitaxial CFS films were characterized by means of cross-sectional transmission electron microscopy (TEM), nanobeam electron diffraction (ED), and $^{57}$Fe conversion electron M\"ossbauer spectroscopy. From these detailed characterizations, we have observed highly ordered $L2_{1}$ structures in the CFS layers\cite{Yamada}. Next, we patterned submicronsized resist mask structures on the CFS films using a conventional electron-beam lithography. Then an Ar ion milling technique is employed to form the wire-shaped CFS spin injector and detector with 300 nm in width. One CFS wire is connected to two square pads to facilitate domain wall nucleation, while the other has pointed-end edges. Using the two different wire shapes, we can control the magnetization configuration by adjusting external magnetic fields ($H$), where $H$ is applied along the CFS wires. Finally, top Cu strips, 200 nm in width and 100 nm in thikness, bridging the CFS wires and bonding pads were patterned by a conventional lift-off technique. Prior to the Cu deposition, the surfaces of the CFS wires were well cleaned by the Ar ion milling with a low accelerating voltage, resulting in low resistive ohmic interfaces with $R_{\rm CFS/Cu}$ $\approx$ 1 ${\rm f}\Omega {\rm m}^2$. Nonlocal and local spin valve measurements were carried out by a conventional current-bias lock-in technique ($\sim 200$ Hz).  The resistivities for the prepared CFS and Cu wires are 90.5 $\mu \Omega$cm and 2.5 $\mu \Omega$ cm at RT and 54.6 $\mu \Omega$cm and 1.2 $\mu \Omega$cm at 80 K, respectively.  

\section*{Acknowledgements}
This work was supported by Fundamental Research Grants from CREST-JST and PRESTO-JST.

\section*{Author Information}
T. K., N. H. and K. H. carried out the device fabrications and the transport measurements.  S. Y., M. M. and K. H. prepared the epitaxial Co$_{2}$FeSi films.  
T. K. and K. H. performed the data analysis.  
T. K., M. M. and K. H. planned the present project and wrote the paper.

%%%%%%%%%%%%%%%%%%%%%%%%%%%%%%%%

%\end{document}

\newpage 

\section*{Supplementary information}
\subsection*{Introducing the resistance area product for nonlocal spin signals}

In the lateral spin valve consisting of two ferromagnetic wire bridged by a nonmagnetic strip as shown in Figs.\ 2a and 2b, the nonlocal spin signal $\Delta R_{\rm S}$ based on a one-dimensional spin diffusion model is given\cite{VFmodel, Takahashi_B} by 
\begin{equation}
\Delta R_{\rm S} =  e^{-\frac{d}{\lambda_{\rm N}}}
\frac{ R_{\rm SN} 
\left( P_{\rm I} \frac{R_{\rm  SI_{inj}}}{R_{\rm SN}} +
	P_{\rm F} \frac{R_{\rm  SF_{ini}}}{R_{\rm SN}} \right)
\left( P_{\rm I} \frac{R_{\rm  SI_{det}}}{R_{\rm SN}} +
	P_{\rm F} \frac{R_{\rm  SF_{det}}}{R_{\rm SN}} \right)}
{\left(1+ 2 \frac{R_{\rm  SI_{inj}}}{R_{\rm SN}} + 2 \frac{R_{\rm  SF_{ini}}}{R_{\rm SN}} 
\right)\left(1+ 2 \frac{R_{\rm  SI_{det}}}{R_{\rm SN}} + 2 \frac{R_{\rm  SF_{det}}}{R_{\rm SN}} 
\right) - e^{-\frac{2d}{\lambda_{\rm N}}}.}
\end{equation}
Here $P_{\rm F}$ and $P_{\rm I}$ are the bulk and interface spin polarizations of the ferromagnetic electrode, respectively, and $R_{\rm SF_{\rm inj}}$, $R_{\rm SF_{\rm det}}$ and $R_{\rm SN}$
are the spin resistances for the ferromagnetic injector, detector and the nonmagnetic strip, respectively. Also, $R_{\rm SI_{\rm inj}}$ and $R_{\rm SI_{\rm det}}$ are the interface spin resistances for the injecting and detecting junctions. $d$ and $\lambda_{\rm N}$ are the separation distance between the injector and detector and the spin diffusion length for the nonmagnetic strip.  
The spin resistance is defined as $2 \rho \lambda /( (1-P^2)S) $, where $\rho$, $\lambda$ and $S$ are the resistivity, the spin diffusion length and the effective cross section for the spin current, respectively. The interface  spin resistance $R_{\rm SI}$ is defined by $2 {\rm RA}/((1-P_{\rm I}^2)S)$, where RA is the resistance area product.  

In the nonmagnetic strip with a long spin diffusion length over a few hundred nanometer, $S$ is given by the cross section of the strip.  
On the other hand, in the ferromagnets with a short spin diffusion length less than 10 nm, $S$ is given by the size of the junction in contact with the nonmagnetic strip because the spin current abruptly decays in the vicinity of the F/N interface\cite{Kimura03}. Moreover, for the ohmic junction, the interface resistance is typically a few hundred mili ohm, which is also much smaller than $R_{\rm SN}$. When $R_{\rm SN} \gg R_{\rm SF}, R_{\rm SI}$, the above equation can be simplified as  
\begin{equation}
\Delta R_{\rm S} \approx 
\frac{
(P_{\rm F} R_{\rm  SF_{inj}} + P_{\rm I} R_{\rm  SI_{inj}})
(P_{\rm F} R_{\rm  SF_{det}} + P_{\rm I} R_{\rm  SI_{det}})
}
{2R_{\rm SN} \sinh \left( {d}/{\lambda_{\rm N}} \right) }.
\end{equation}
It should be noted that although $R_{\rm SF}$ increases with $P$ and diverges at $P = 1$, 
the condition of $R_{\rm SF}$ $\ll$ $R_{\rm SN}$ is still valid for $P < 0.9$ for the CFS film.

By introducing the junction sizes $(S_{\rm inj}$, $S_{\rm det}$ and $S_{\rm N})$, this equation can be revised as 
\begin{equation}
\Delta R_{\rm S} \approx \frac{S_{\rm N}}{S_{\rm inj}S_{\rm det}}
\frac{ \left( \frac{ P_F}{(1-P_F^2)} \rho_{\rm F} \lambda_{\rm F} 
+ \frac{P_I}{(1-P_I^2)} {\rm RA}_{\rm F/N} \right)^2 }
{\rho_{\rm N} \lambda_{\rm N} \sinh \left( {d}/{\lambda_{\rm N}} \right)}.
\end{equation}

By defining $\Delta R_{\rm S} A$ as $\Delta R_{\rm S} (S_{\rm inj} S_{\rm det}/{S_{\rm N}})$, we obtain 
\begin{equation}
\Delta R_{\rm S} A \approx 
\frac{ \left( \frac{ P_F}{(1-P_F^2)} \rho_{\rm F} \lambda_{\rm F} 
+ \frac{P_I}{(1-P_I^2)} {\rm RA}_{\rm F/N} \right)^2 }
{\rho_{\rm N} \lambda_{\rm N} \sinh \left( {d}/{\lambda_{\rm N}} \right)}.
\end{equation}
In Eq. (4), the influence of the junction sizes on the spin signal can be normalized.  Thus, $\Delta R_{\rm S} A$ allows us to fairly evaluate the generation efficiency of the pure spin current 
for various combinations of a ferromagnetic metal and a nonmagnetic one.

\subsection*{Estimation of interface resistance }
The interface resistance was estimated by measuring the 4-terminal resistances with local spin valve configuration. In this configuration, the total resistance consists of the resistance of the Cu wire and the two interface resistances.  Since the resistance of the Cu wire can be estimated from the resistivity for Cu, we can roughly calculate the interface resistance by subtracting the resistance for the Cu wire from that in the local spin valve configuration.  By using the relation that the difference in the resistance $\Delta R$ is given by ${\rm RA}_{\rm F/N}(S_{\rm inj} + S_{\rm det})$, we can obtain ${\rm RA}_{\rm F/N}$.  
For the CFS/Cu LSVs,  $\Delta R$ was $\sim$ 30 m$\Omega$, 
indicating ${\rm RA_{CFS/Cu}}$ $\sim$ 1 f$\Omega$m$^2$.

\subsection*{Estimation of spin polarization}
By fitting the experimental data on the distance dependences of the $\Delta R_{\rm S}A$ using Eq. (4), we can estimate the spin polarization of the spin injector 
in the LSV systems.   For the Py/Cu LSVs, $R_{\rm Py/Cu}$ is less than 0.1 ${\rm f} \Omega {\rm m}^2$, much smaller than $\rho_{\rm Py} \lambda_{\rm Py}$ (0.75 ${\rm f} \Omega {\rm m}^2$). Thus, we can neglect the second term in the numerator of Eq. (3), then obtain $P_{\rm Py} \sim$ 0.3 at RT and 0.35 at 80 K, respectively with assuming $\lambda_{\rm Py, RT} =$ 3 nm and $\lambda_{\rm Py, 80 K} = 5$ nm\cite{Kimura01}.

For the CFS/Cu LSVs, as described in the previous section, 
$R_{\rm CFS/Cu}$ can be approximately estimated as $\sim$1 ${\rm f} \Omega {\rm m}^2$.  
Because of the following reasons, 
we assumed that the spin diffusion length for CFS ($\lambda_{\rm CFS}$) is the same 
order of that for CFSA ($\lambda_{\rm CFSA}$), which was reported as 2.2 nm at RT and 3 nm at 14 K 
in recent study of the vertical magnetoresitance device\cite{Nakatani2_B, Taniguchi}.  
The spin-diffusion length is proportional to the magnitude of the spin-orbit interactions. 
Since the atomic number of Al is close to Si, we expect that the magnitude of spin-orbit interaction in CFS should be almost same order of CFSA.  
From these considerations, we expect that $\lambda_{\rm CFS}$ is 
the same level or shorter than $\lambda_{\rm CFSA}$.   
Since the use of the longer $\lambda_{\rm CFS}$ prevents an overestimation of $P_{\rm CFS}$, 
we use $\lambda_{\rm CFS, RT} = 2 \sim 4$ nm and $\lambda_{\rm CFS, 80 K} = 3$ $\sim$ 6  nm, 
where the minimum value corresponds to $\lambda_{\rm CFSA}$.  
We then obtained $P_{\rm CFS} = 0.50 \sim 0.71$ at RT and $P_{\rm CFS} = 0.66 \sim 0.81$ at 80 K 
with assuming $P_{I} = 0.5 \sim 0.7 $, which is typical interface spin polarization between the Co-based alloy and Cu \cite{CPP1,CPP2}.

%Recent study of the vertical magnetoresitance device composed of Co$_{\rm 2}$FeSiAl (CFSA) claims that the spin diffusion length for CFSA is 2.2 nm at RT and 3 nm at 14 K\cite{Nakatani2, Taniguchi}.  

\subsection*{
Comparison of the device performance between LSVs 
with metallic and resistive interface resistances
}

In the present paper, we discuss the LSVs only with metallic junctions, since Eq. (1) is valid only for the condition of 
$\rho_{\rm F} \lambda_{\rm F}$, ${\rm RA_{F/N}} \ll \rho_{\rm N} \lambda_{\rm N}$.  
When the above condition is not satisfied, for example, 
the LSVs with the resistive interface 
(${\rm RA_{F/N}} \gg  \rho_{\rm N} \lambda_{\rm N}$) \cite{Wang, Fukuma, Hoffmann}, 
the device performance cannot be evaluated by Eq (1). 
In order to fairly compare the device performance of different type LSVs, 
one should focus on the injection efficiency of the pure spin current.  
For the metallic junctions, the injection efficiency $\eta_{I_S}$ of the pure spin current, 
which is defined by the ratio of the spin current $I_S$ injected 
into the ferromagnetic contact to the excited charge current $I_C$, can be calculated as 
\begin{equation}
\eta_{I_S} \equiv \frac{I_S}{I_C} \approx 
\frac{1}{2} \frac{S_{\rm N}}{S_{\rm Inj}}
\frac{ \left( \frac{ P_F}{(1-P_F^2)} \rho_{\rm F} \lambda_{\rm F} 
+ \frac{P_I}{(1-P_I^2)} {\rm RA}_{\rm F/N} \right)}
{\rho_{\rm N} \lambda_{\rm N} \sinh \left( {d}/{\lambda_{\rm N}} \right)}.
\end{equation}
For example, the injection efficiency of the present CFS/Cu LSV is estimated to 
be $\sim 0.5$ with $d = 100$ nm.

On the other hand, the efficiency for the LSV with the resistive interface, 
where ${\rm RA_{F/N}} \gg  \rho_{\rm N} \lambda_{\rm N}$, is given by
\begin{equation}
\eta_{I_S} \approx 
\frac{1}{4} P_{\rm I} \frac{S_{\rm Det}}{S_{\rm N}}
\frac{\rho_{\rm N} \lambda_{\rm N}}{{\rm RA_{\rm F/N}}}
\exp^{-\frac{d}{\lambda_{\rm N}}}.
\end{equation}
In this case, the efficiency decreases with increasing 
the interface resistance area product $\rm RA_{\rm F/N}$. 
For example, for the Py/Ag LSVs with moderate interface resistances, 
where the large spin signals comparable to the present CFS/Cu LSV 
have been reported,\cite{Fukuma, Hoffmann} 
the interface resistance area product $\rm RA_{\rm F/N}$ is 100 f$\Omega$m$^2$. 
Thus, the injection efficiency is estimated to be $\sim$ 
0.005 with $d =$ 100 nm.

%\subsection*{Estimation of power consumption}
%The electrical power for switching the magnetization by 
%the pure spin current injection in the CFS/Cu LSV can be roughly estimated as follows.  
%The magnetization of the ferromagnetic nanodot can be switched by injecting a 
%spin current of $\sim$ 100  $\mu$A \cite{Ikeda}.  Since the injection efficiency of 
%the pure spin current is 0.5 in the present CFS/Cu LSV, 
%the spin current can be injected nonlocally by flowing a current of $\sim$ 200 $\mu$A 
%in the current probe with the two-terminal resistance of $\sim$10 $\Omega$.  
%This enables us to switch the magnetization with an electric power less 
%than one microwatt.  

%\clearpage
%\noindent{{\bf Figure Legends}}

%{\small\noindent Figure 1: {\bf }.
%\vspace{8mm}

%\noindent Figure 2: {\bf }.
%\vspace{8mm}

%\noindent Figure 3: {\bf }.
%\vspace{8mm}

%\noindent Figure 4: {{\bf }.

%\clearpage
%\begin{figure}
%\includegraphics[width=8cm]{Fig1.eps}
%\caption{ }
%\end{figure} 

%\end{document}

\end{document}